\documentclass[twocolumn,showpacs,preprintnumbers,amsmath,amssymb]{revtex4}
\usepackage{graphicx}
\usepackage{dcolumn}
\usepackage{bm}

\begin{document}

\title{Fermi Velocity Spectrum and Incipient Magnetism in TiBe$_{2}$}

\author{T. Jeong, A. Kyker}
 
\author{W.E. Pickett}
 
\affiliation{
Department of Physics, University of California, Davis, California 95616
}

\date{\today}

\begin{abstract}
We address the origin of the incipient magnetism in TiBe$_2$
through precise first principles calculations, which overestimate the
ferromagnetic tendency and therefore require correction to account for
spin fluctuations.  TiBe$_2$ has sharp fine structure in its
electronic density of states, with a van Hove singularity only
3 meV above the Fermi level.  Similarly to the isovalent weak ferromagnet 
ZrZn$_2$, it is flat bands along the K-W-U lines of hexagonal face of
the fcc Brillouin zone make the system prone to magnetism, and more so if 
electrons are added.  We find that the Moriya $B$ coefficient
(multiplying $\frac{\omega}{q}$ in the fluctuation susceptibility
$\Delta \chi(q,\omega)$) is divergent
when the velocity vanishes at a point on the Fermi surface, which is
very close (3 meV) to occurring in TiBe$_2$.  In
exploring how the FM instability (the $q$=0 Stoner enhancement is 
$S\approx 60$) might be suppressed by fluctuations 
in TiBe$_2$, we calculate that the Moriya A coefficient (of $q^2$) is negative,
so $q$=0 is not the primary instability.  Explicit calculation of
$\chi_o(q)$ shows that its 
maximum occurs at the X point $(1,0,0)\frac{2\pi}{a}$;  
TiBe$_2$ is thus an incipient {\it anti}ferromagnet rather than 
ferromagnet as has been supposed.
We further show that simple temperature
smearing of the peak accounts for most of the temperature dependence
of the susceptibility, which previously had been attributed to
local moments (via a Curie-Weiss fit), and that energy dependence of the
density of states also strongly affects the magnetic field variation
of $\chi$.
\end{abstract}

\pacs{Valid PACS appear here}

\maketitle

\section{\label{sec:level1}Introduction}

The cubic Laves compound TiBe$_{2}$ was already shown forty years ago
to have quite unusual behavior of the magnetic 
susceptibility $\chi(T)$ and the Knight shift.\cite{saji}
$\chi^{-1}$ showed a strong increase with lowering temperature 
but a clear deviation
from Curie-Weiss form, while
the Knight shift was temperature dependent and negative.
The magnetic properties of TiBe$_{2}$ have been controversial since 
Matthias  {\it et al.}\cite{matthias} interpreted the susceptibility
peak at 10 K in TiBe$_{2}$ as 
itinerant antiferromagnetism (AFM) with an associated
magnetic moment of 1.64$\mu_{B}$, and Stewart {\it et al.} reported a
transition at 2 K that seemed characteristic of magnetic ordering.

However, a clear picture has emerged gradually after the idea 
of weak itinerant antiferromagnetism had been abandoned because of the 
subsequent lack of experimental evidence\cite{rak, takagi}.
Many experiments have shown that TiBe$_{2}$ is instead a strongly 
enhanced paramagnet \cite{acker,monod,shaltiel}
and undergoes a metamagnetic transition\cite{rhodes,wohlfarth,acker2} 
(field-driven ferromagnetism)
around 5.5 T. 
Also one can see similarity to the magnetic behavior of  
Ni$_{3}$Ga by comparing the values of the low temperature 
susceptibility, $\chi=1.65 \times 10^{-2}$ emu/mole for 
Ni$_{3}$Ga\cite{schinkel} and  
$\chi=0.90 \times 10^{-2}$ emu/mole for 
TiBe$_{2}$\cite{matthias}.   
Based on the magnetization data of Monod {\it et al}\cite{monod}
Wohlfarth\cite{wohlfarth} suggested the transition at 5.5 T should
be first order. 
Wohlfarth's considerations received at least partial support from 
theoretical band-structure considerations coupled with the de Haas-van
Alphen data of van Deursen {\it et al}\cite{deursen}.

Clarity began to arise with the extensive experiments of Acker 
{\it  et al.} who interpreted their
magnetization data\cite{acker}
in fields to 21T and the magnetization data of Monod {\it et al.}
\cite{monod} as evidence 
for exchange-enhanced paramagnetism or spin fluctuations in TiBe$_{2}$.
They found the system TiBe$_{2-x}$Cu$_{x}$ to become FM at a critical 
concentration $x_{cr}=0.155$.
Stewart {\it et al.}\cite{stewart} measured the specific heat of   
TiBe$_{2}$ ($\gamma =42~$mJ/mole K$^{2}$) 
at low temperature in 0 and 7T and interpreted the behavior as 
evidence of spin fluctuations. 

The isoelectronic isostructural material ZrZn$_{2}$ is considered 
a classic example of an weak itinerant ferromagnet.
Magnetic measurements find
very small magnetic moments (values from
0.12 to 0.23 $\mu_{B}$ )\cite{ZrZn1, ZrZn2}, hence the characterization
as a weak ferromagnet.
The magnetization of ZrZn$_2$ increases substantially with field,  but 
unlike TiBe$_2$ with its metamagnetic transition, the increase continues
smoothly to fields as high as 35 T.
The Curie temperature T$_{C}$ drops approximately linearly
with pressure, from 29 K at $P=0$
to 4K at $P=16$ kbar,
which extrapolates to a quantum critical point (QCP) at
$P=18-20$ kbar.
The report of superconductivity coexisting with ferromagnetism 
in ZrZn$_{2}$ near this QCP\cite{Pfleiderer}
enlivened both theoretical and experimental 
attention, but more recently it has been shown\cite{surfaceSC} 
there is no bulk superconductivity.
TiBe$_2$, on the other hand, has been nearly
addressed only rarely for the past twenty years.

The complex temperature-field behavior of TiBe$_2$ has led to many
speculations about the microscopic mechanisms.  Of course spin
fluctuations play a central part, and the highly enhanced susceptibility
suggests this system is near a quantum critical point (at slightly enlarged 
lattice constant, say, as well as for the Cu alloying).  
If FM fluctuations dominate, then a metamagnetic
transition (field-driven FM state) around 5 T would make sense.  If
AFM fluctuations dominate, application of a field suppresses the 
fluctuations, providing another way to interpret specific heat under
applied field.\cite{stewart2}  The anomalies in the conduction 
electron spin resonance (CESR) linewidth\cite{ioshpe} around 2 K
have been interpreted in
terms of a thermal spontaneous magnetism,\cite{theory} and a
decrease in the resistivity is also seen at that 
temperature.\cite{acker}  All of these scenarios are sensitive 
to the Fermi surface shape, velocity spectrum, and possibly the
energy dependence of the density of states near the Fermi energy,
and it is these questions that we address in this paper.

Band structure intricacies by themselves also can come into play.
Shimizu showed\cite{theory} that an independent electron
system with magnetic coupling can undergo a first-order 
transition to a ``spontaneous
thermal magnetism'' state (within a range  $T_1 < T < T_2$)
if it is highly enhanced and if the Fermi level lies within a
local minimum in the density of states.  The effects of magnetic
fluctuations should of course be added\cite{moriyaueda} to the
free energy of both the ordered and disordered phases to make
this treatment more realistic.

Local density approximation (LDA) energy band studies of 
TiBe$_{2}$ have been reported previously
\cite{freeman1,freeman2,groot}.
Those studies revealed a split narrow peak in in the density of states
(DOS) N(E) near the Fermi energy (E$_F$), with 
calculated Stoner factors $I N(E_F)$ greater 
than unity, giving the Stoner instability to FM.  Here $I$ is the 
Stoner exchange interaction averaged over the
Fermi surface.  Thus, as for a few cases that have come to light more
recently,\cite{djs1,djs2} ferromagnetism is incorrectly 
predicted, indicating the need to
account for magnetic fluctuations not included in LDA that will suppress
magnetic ordering.  By comparing the calculated value of N(E$_F$) with the
measured susceptibility, a Stoner enhancement
S = [1 - $I N(E_F)$]$^{-1} \approx$ 60 was
obtained, making TiBe$_2$ a more strongly exchange enhanced metal than
Pd.

All of these calculations, carried out 25 years ago,
used shape approximations for the density and potential, 
and for a detailed investigation of the weak ferromagnetism precise
electronic structure methods are required.
In this work,
the precise self-consistent full potential 
linearized-augmented-plane-wave (FLAPW) method and full potential 
local orbital minimum basis band structure scheme (FPLO) are 
employed to investigate thoroughly the electronic and magnetic 
properties of TiBe$_{2}$ based on the density functional theory.
We compared and checked the calculation results of the both methods.
We consider the effect of magnetism on the band structure and Fermi
surface, Fermi velocity and compare with experiment and previous 
band calculations.

\section{Crystal Structure}

TiBe$_{2}$ crystallizes into a cubic Laves phase C15 crystal structure.
The C15 (AB$_{2}$ ) structure is a close packed structure 
and the site symmetry is high for the two constituents. 
Ti atoms occupy the positions of a 
diamond sublattice while the Be atoms form a network of interconnected 
tetrahedra, with two formula units per cell. 
Since the major contributions to $N(E_{F})$ come from 
Ti, the local environment of Ti atoms is particularly important 
to keep in mind.
Each Ti is surrounded by 12 Be neighbors at a distance of 
2.66  $\AA$
and tetrahedrally by four Ti neighbors a distance 2.78 $\AA$ away. 
The TiBe$_2$ structure belongs to the Fd3m
space group with Ti occupying the $8a$ site, and Be the $16d$ site.
The site symmetry of Ti is $\bar{4}3m$(tetrahedral) 
and Be has $\bar{3}m$ site symmetry.  The atomic positions are symmetry
determined, and
we used experimental lattice constant 6.426~$\AA$ for all calculations.

\section{Method of Calculations}

We have applied the full-potential 
nonorthogonal local-orbital minimum-basis (FPLO) scheme within the local 
density approximation (LDA).\cite{koepernik}
In these scalar relativistic calculations we 
used the exchange and correlation potential of Perdew and Wang.\cite{perdew}
Ti $3s,3p,4s,4p,3d$ states and Be $2s,2p,3d$ were included as 
valence states. All lower states were treated as core states.
We included the relatively extended semicore $3s,3p$ states of Ti 
as band states because of the considerable overlap of these 
states on nearest neighbors.
This overlap would be otherwise neglected in our FPLO scheme. Be 3d states
were added to increase the quality of the basis set. The spatial extension
Of the basis orbitals, controlled by a confining potential $(r/r_{0})^4$, was 
optimized to minimize the total energy. 

The self-consistent potentials were 
carried out on a mesh of 50 k points in each direction of the Brillouin zone,
which corresponds to 3107 k points in the irreducible zone.
A careful sampling of the Brillouin zone is necessary to account carefully 
for the fine structures in the density of states near Fermi level E$_{F}$.
For the more delicate numerical integrations, band energies were extracted from
FPLO in an effective mesh of 360 k points in each direction.
A separate tool was developed to extract energy isosurfaces
with gradients from the scaler energy grid. The isosurfaces
were then used to calculate density of states and velocity moments.

To check carefully the fine structure that we will discuss,
we also repeated several calculations with the general potential 
linearized augmented plane wave (LAPW)
method,\cite{singh} as implemented in the WIEN2K code.\cite{wien}
Relativistic effects were included at the scalar relativistic level. 
However, we verified that the magnetic moment with the 
experimental structure is not sensitive to the inclusion of the 
spin-orbit interaction.
For the generalized gradient approximation (GGA) calculations, we used 
the exchange-correlation functional of Perdew, Burke, and Ernzerhof. 
\cite{burke}
We choose the muffin-tin spheres $R_{MT}=2.6$ a.u. for Ti, 
$R_{MT}=2.1$ a.u. for Be and 
a basis set determined by a plane-wave cutoff of $R_{MT}K_{max}=7.0$,
which gives good convergence. The Brillouin zone samplings 
were done using the special k point method with 1280 points
in the irreducible zone. 

\section{Results and Discussions}

\begin{figure}
\includegraphics[height=8.5cm,width=8.5cm,angle=-90]{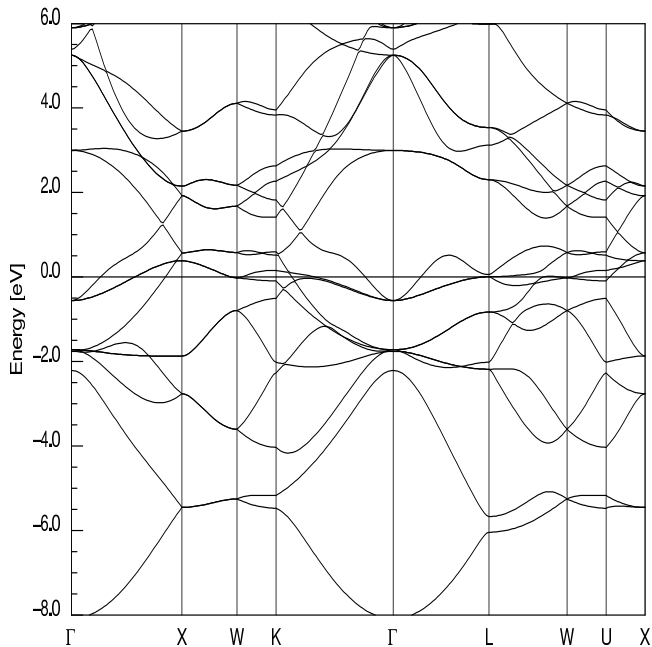}
\caption{The full LDA band structure of non-magnetic TiBe$_{2}$ along
symmetry lines showing that there are several bands near the Fermi level
(taken as the zero of energy) with weak dispersion; they are 
primarily Ti $3d$ in character.} 
\label{band}
\end{figure}

\begin{figure}
\includegraphics[height=8.5cm,width=8.5cm,angle=-90]{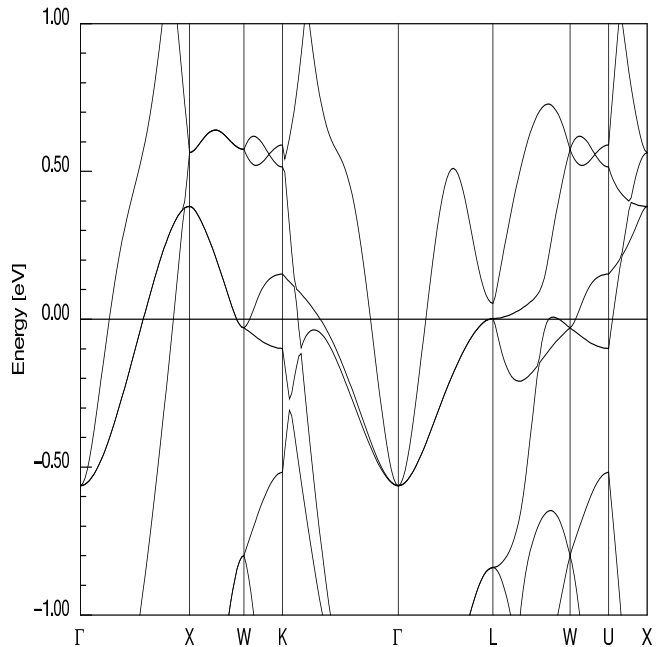}
\caption{Band structure of non-magnetic TiBe$_{2}$ of Fig. \ref{band}
on an expanded scale
near Fermi level.
The flat bands along L-W-U/K-L lines (the hexagonal face of the fcc
Brillouin zone) give rise to the density of states structure discussed
in the text.} 
\label{blowup}
\end{figure}

\begin{figure}
\includegraphics[height=8cm,width=8cm]{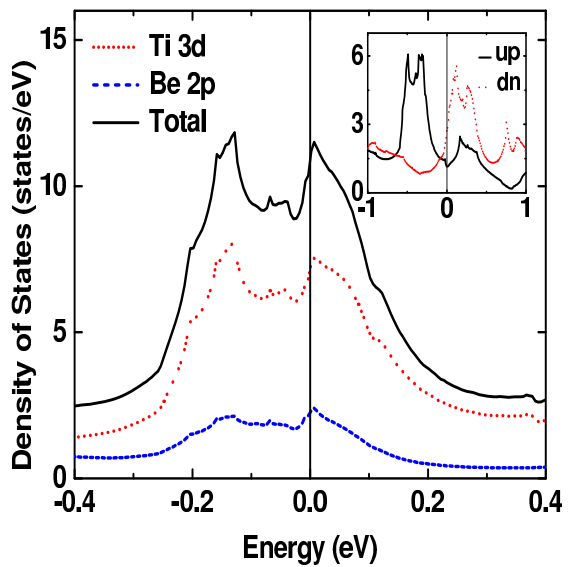}
\caption{The total and atom-projected density of 
states (Ti, short dashed line; Be, the lower, long dashed line)
for non-magnetic TiBe$_{2}$ per primitive cell. 
The inset gives the density of states for the ferromagnetic TiBe$_{2}$ showing 
the exchange splitting 0.6 eV. The peak of the DOS 
for the majority spin is entirely 
below the Fermi level while that of the minority spin is above 
the Fermi level.} 
\label{dos}
\end{figure} 

\begin{figure}
\includegraphics[height=5.6cm,angle=-0]{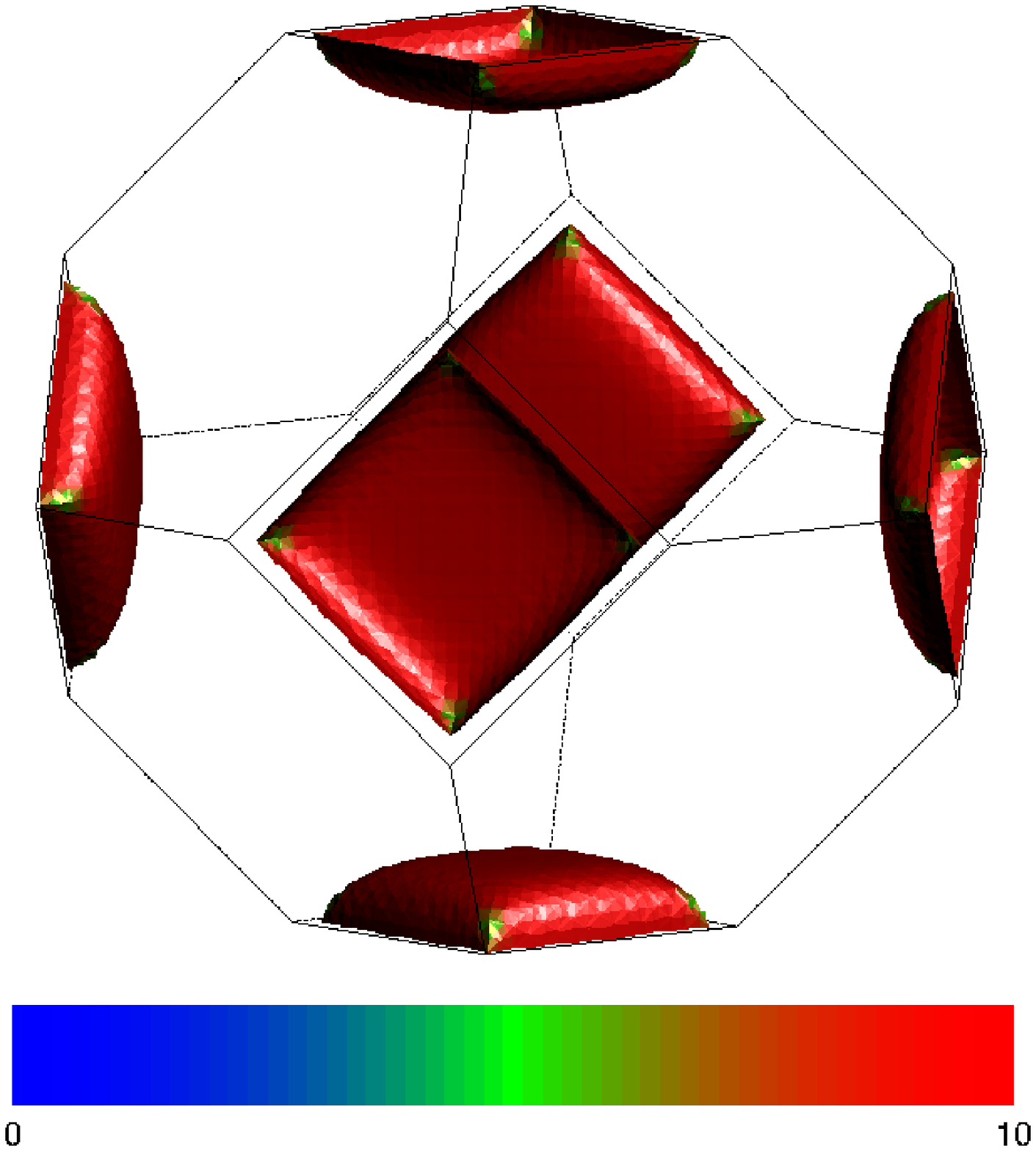}
\includegraphics[height=5.6cm,angle=-0]{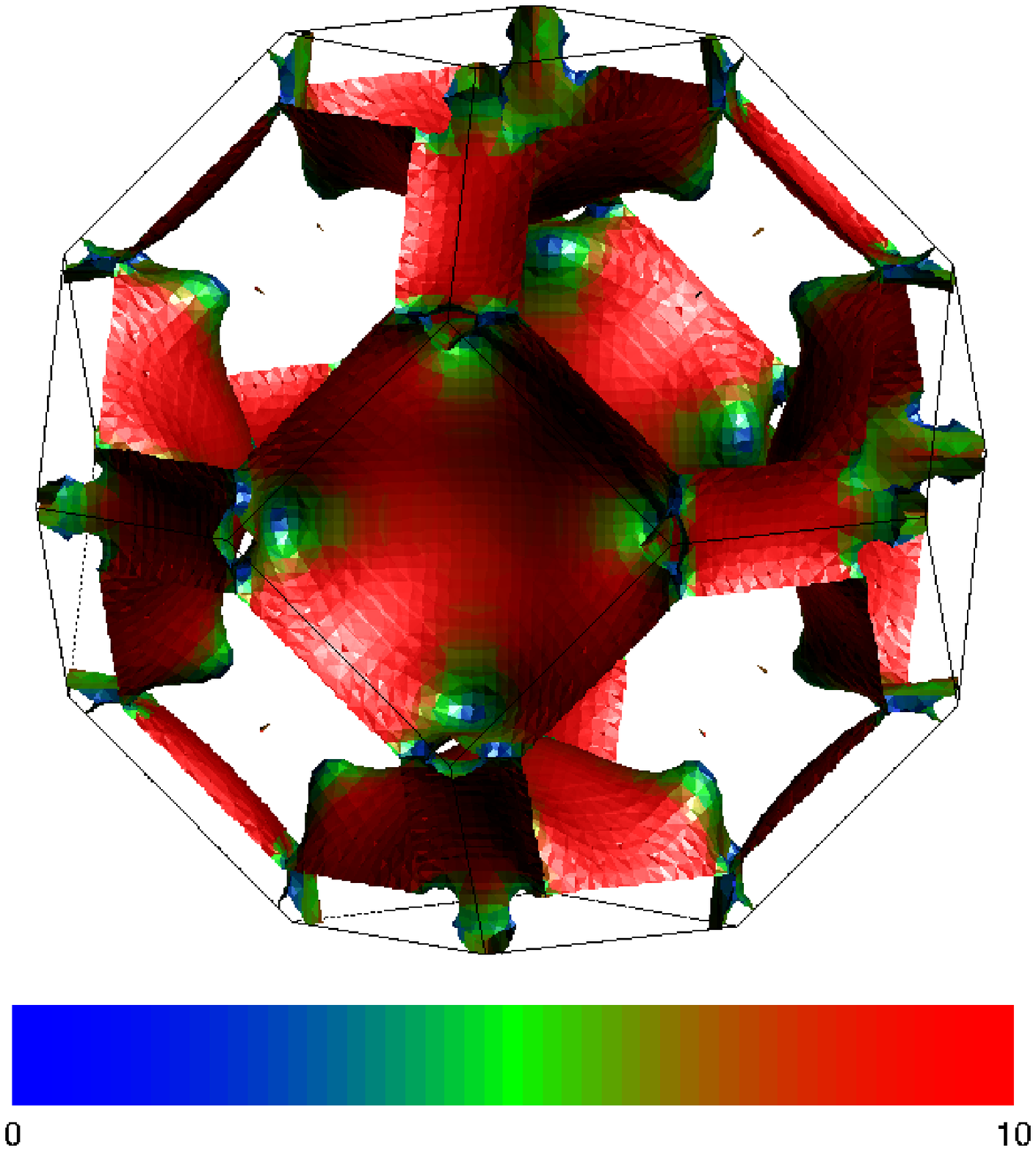}
\includegraphics[height=5.6cm,angle=-0]{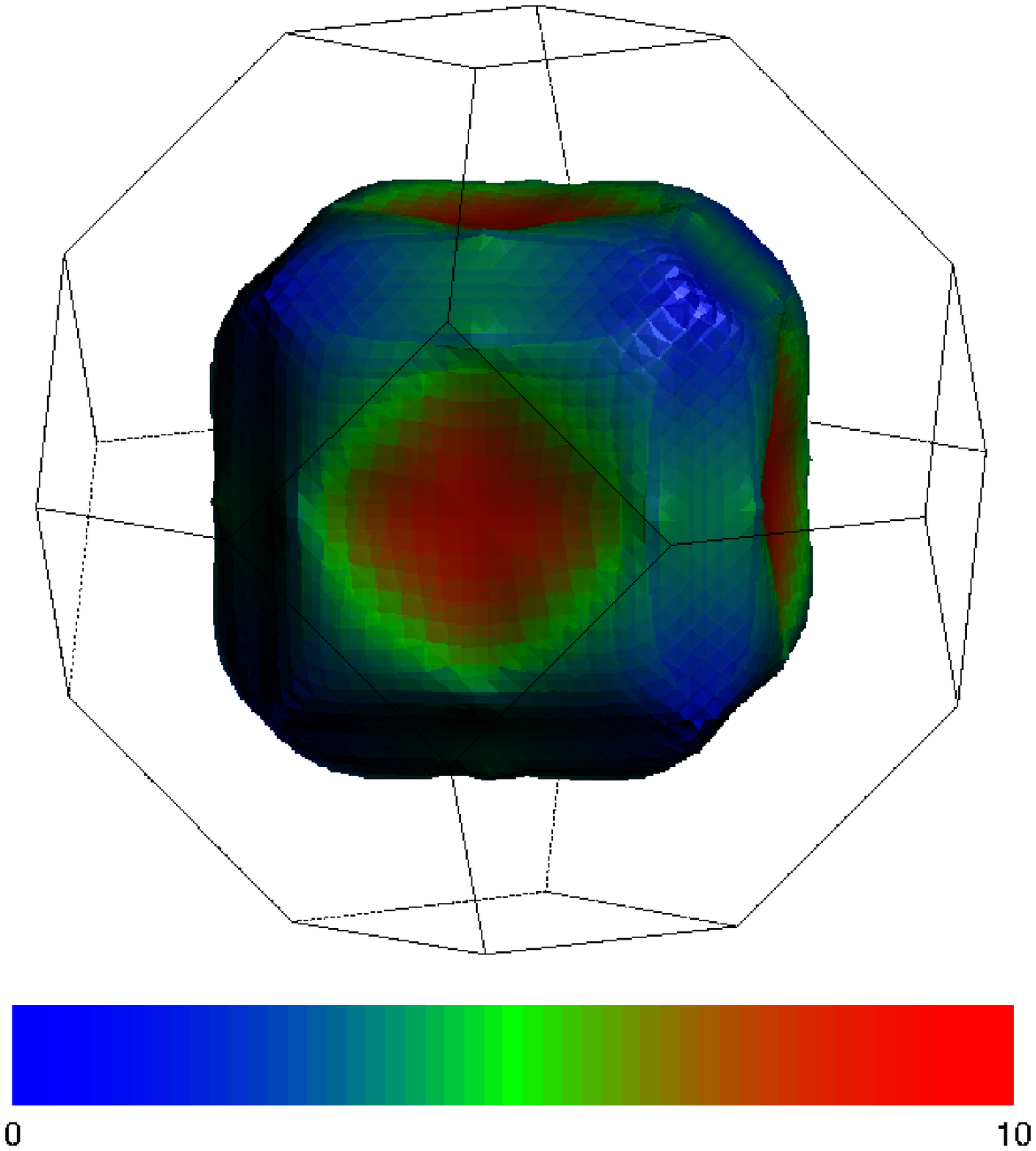}
\includegraphics[height=5.6cm,angle=-0]{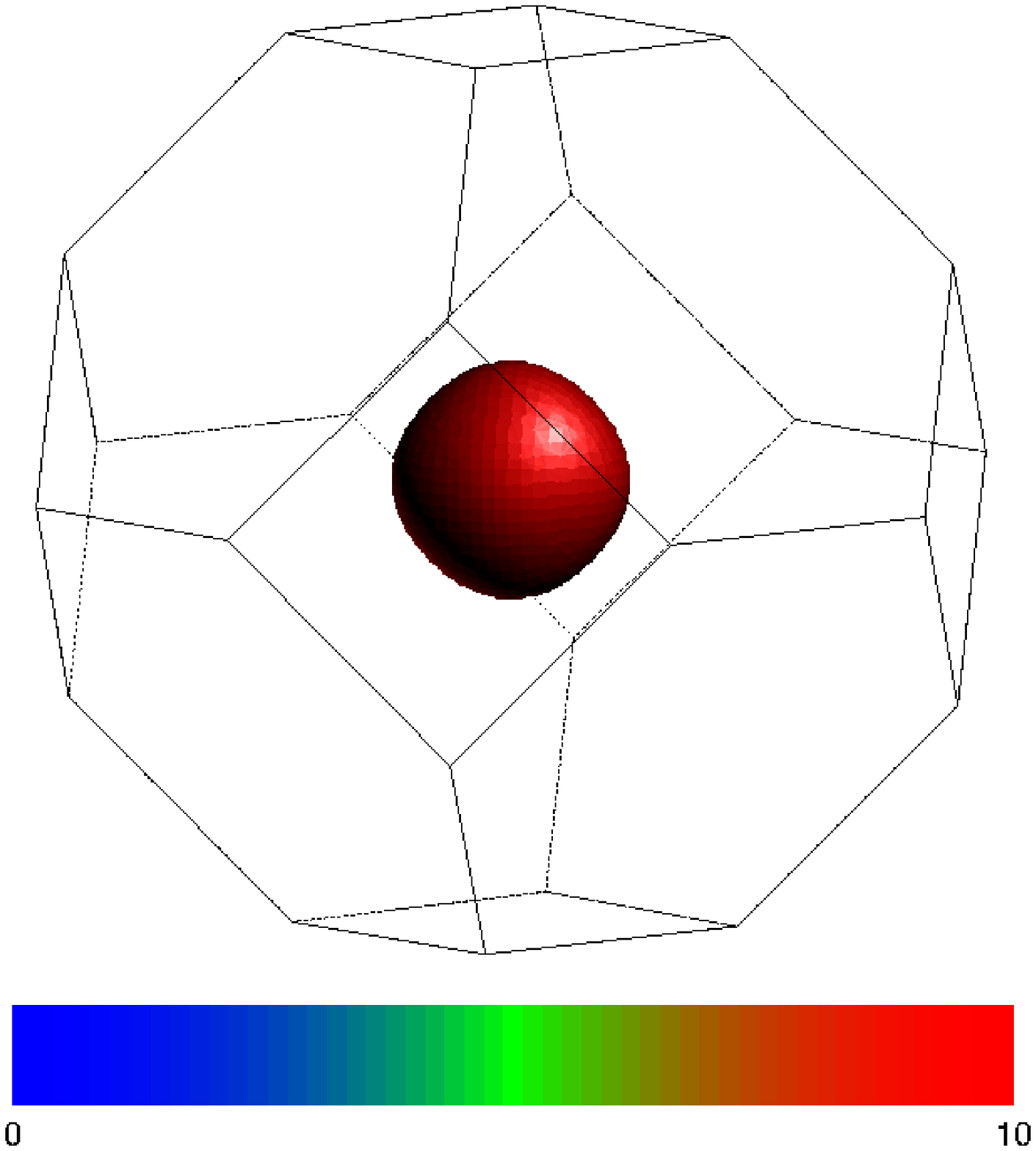}
\caption{(Color online)
Fermi surfaces, from top: band 14, X-centered pillows; band 15, 
primarily X-centered jungle gym; band 16, $\Gamma$-centered 
pseudocube; band 17, $\Gamma$-centered sphere.  
Fermi velocities colored dark (red) for lowest to lighter (blue) for
highest.  Magnitudes of velocities are discussed in Sec. IV.A.}
\label{Fs}
\end{figure}

For orientation we first show the full nonmagnetic
band structure of TiBe$_{2}$ in Fig. \ref{band},
which is consistent with earlier calculations of 
\cite{freeman1,freeman2, groot}.
The Be $2s$ bands lie between -8 eV and -2 eV. 
Above them the bands are of mixed $s, p$ character, centered on the Be 
as well as the Ti site.
Near the Fermi level there are several bands with 
weak dispersion, being of primarily Ti 
$3d$ character. 
The bands at K and L are hybridized strongly, while at X the $s,p$ character
is the main character.  As noted also by Jarlborg and 
Freeman,\cite{freeman1,freeman2} one band at L falls extremely close
to E$_F$ (3 meV below).  This band is doubly degenerate along $\Gamma$-L,
and the L point forms the maximum of band 15 and a saddle point for  
band 16.  As the Fermi energy rises (for added electrons, say) the
Fermi surface sweeps through the L point saddle, where the band has
a vanishing velocity by symmetry.  This vanishing velocity is
discussed below.  There is another doubly degenerate band very near
E$_f$ at the W point.

The density of states (DOS) is shown near E$_F$ in Fig. \ref{dos}.
The Fermi energy $E_{F}$ falls extremely close to  
the edge of a very narrow peak in the DOS.
The DOS peak arises from Ti $d$ bands hybridized with Be $p$ states.
Flat bands close to Fermi level centered mostly in regions near the 
L-W-U and W-K directions, {\it i.e.} the hexagonal faces of the Brillouin
zone, cause the sharp peak.
Stewart {\it et al.}\cite{stewart} measured the linear specific heat
coefficient for TiBe$_2$ of $\gamma$=42 mJ/K$^{2}$ mole-formula unit. 
The calculated value of $N(E_{F})$=5.33 states/eV/f.u. for TiBe$_2$
corresponds to
a bare value $\gamma_o$=12.6 mJ/K$^{2}$ mole(formula unit), leading
to a thermal mass enhancement 1+$\lambda$=3.3, or $\lambda$=2.3 
arising from phonons, magnetic fluctuations, and Coulomb interactions.

Density functional calculations are usually reliable in calculating 
the instability to ferromagnetism.
The enhanced susceptibility\cite{janak} is given by 
\begin{eqnarray}
\chi(T) =\frac{\chi_{0}}{1-N(E_{F})I} \equiv S\chi_{0}.
\label{susceptibility}
\end{eqnarray}
where $\chi_{0}=\mu_{B}^{2}N(E_{F})$ is the bare
susceptibility obtained directly from the  band
structure and $I$ is the Stoner exchange interaction constant.
Here N(E$_F$) refers to both spins, and hence forward we quote
susceptibility in units where $\mu_B \equiv 1$.
The calculation of $I$ is 
from fixed spin moment calculations\cite{mohn},
in which 
the energy $E(m)$ is calculated subject to the moment 
being constrained to be $m$. The behavior at small $m$ 
is $E(m)=(1/2)\chi^{-1}m^{2} $ from which $I=0.22$ eV can be 
extracted from Eq. \ref{susceptibility}.
This value of $I$ gives $IN(E_{F})=1.2$, larger than unity and
very close to that calculated earlier,\cite{freeman2} corresponding 
to a Stoner ferromagnetic instability.

As for a few other compounds, TiBe$_{2}$ is incorrectly predicted
by LDA to be ferromagnetic.
Since spin-orbit coupling is small in $3d$ magnets, we neglect it, so
the direction of magnetic polarization is not coupled to the lattice.
We have calculated a consistent magnetic moment for TiBe$_{2}$:
$0.97\mu_{B}$/f.u.(FPLO, LDA),
$1.00\mu_{B}$/f.u.(LAPW, LDA),
$1.10\mu_{B}$/f.u.(LAPW, GGA).
This value is considerably larger than an earlier 
calculation\cite{freeman1} (which also reported a much smaller
value for ZrZn$_2$ than obtained from more recent 
calculations\cite{mazin}).
We address the overestimate of the tendency to magnetism below.

\subsection{Fermi Surface and Fermi Velocity}

In Fig. \ref{Fs} we show the nonmagnetic Fermi surfaces shaded by the
Fermi velocities.
The position of $E_{F}$ near L and W points sensitively determine 
the exact shape of some Fermi surfaces.  The shapes can be characterized
as (a) small $\Gamma$-centered electron sphere from band 17, (b) large
$\Gamma$-centered electron pseudocube from band 16, (c) multiply
connected surface mostly enclosing holes around the X point from 
band 15, which we refer to as the jungle gym, and (d) flat hole pillows
centered at each of the three X points.  The doubly degenerate bands
crossing E$_F$ along $\Gamma$-X and X-W guarantee touching of certain
surfaces along these lines.

\begin{figure}
\vskip 5mm
\includegraphics[height=8.5cm,width=8.5cm,angle=-90]{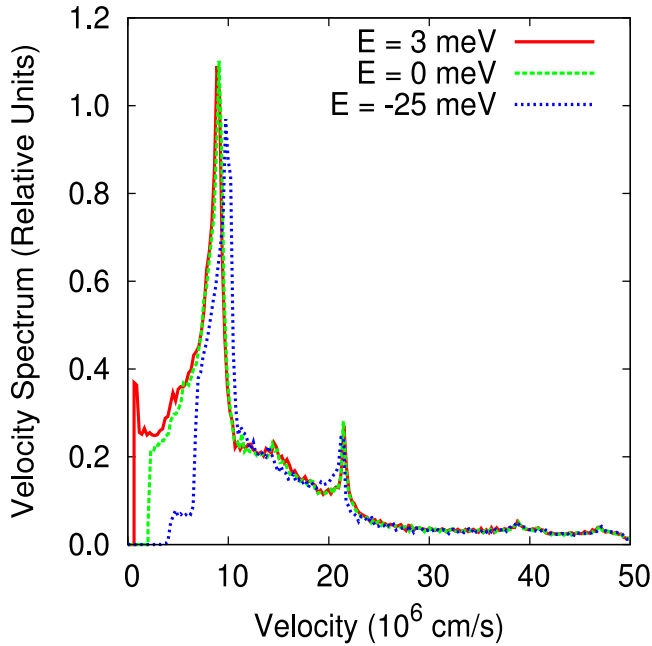}
\caption{(Color online) Fermi velocity spectrum of TiBe$_{2}$. 
The low Fermi velocity states are the primary source of changes to
the density of states.  } 
\label{fv}
\end{figure}

The DOS peak at and above E$_F$ is due to the band
near the L point where the cube-shaped surfaces are about
to form bridging necks. Figure \ref{fv} shows 
how the Fermi velocity spectrum $(N(V;E))$ changes with energy at the peak
just above $E_F$, at $E_F$, and at the first minimum below $E_F$. 
The Fermi velocity spectrum is defined as
\begin{eqnarray}
N(V;E) &=& \sum_{\vec{k}} \delta(E_{\vec{k}} - E) 
\delta(V_{\vec{k}} - V) \\ \nonumber
 &=& \int_{{\cal L}(V;E)}\frac{d{\cal L}_k}{|\vec v_k \times
  \nabla_k |\vec v_k| |},
\end{eqnarray}
with normalization $\int N(V;E) dV = N(E)$.  Here ${\cal L}(V;E)$ is
the line of intersection of the constant energy $E_k = E$ surface with the
constant velocity surface $|\vec v_k| = V$.  The gradient of the velocity in
the denominator makes this distribution delicate to calculate accurately.
$N(E,V)$ was calculated
numerically by extracting a triangulated energy isosurface from the
band structure, then obtaining a velocity histogram of the states
associated with the isosurface.

The spectrum in Fig. \ref{fv} shows, at E$_F$, velocities extending
down to the very low value of 2$\times 10^6$ cm/s, and up to 
5$\times 10^7$ cm/s, a variation of a factor of 25.  Roughly half of the
weight lies below $10^7$ cm/s.  At the van Hove singularity at +3 meV,
the only noticeable difference is additional velocities extending down
to zero due to the vanishing velocity at L (we have not worried about
reproducing the $V\rightarrow 0$ behavior precisely).  At -25 meV,
which is just below the narrow peak at E$_F$, the strong weight in
the spectrum appears only at 7$\times 10^6$ cm/s.   Note that there is
very little change in the high velocity spectrum over small changes
in energy.

\begin{figure}
\includegraphics[height=8.5cm,angle=-90]{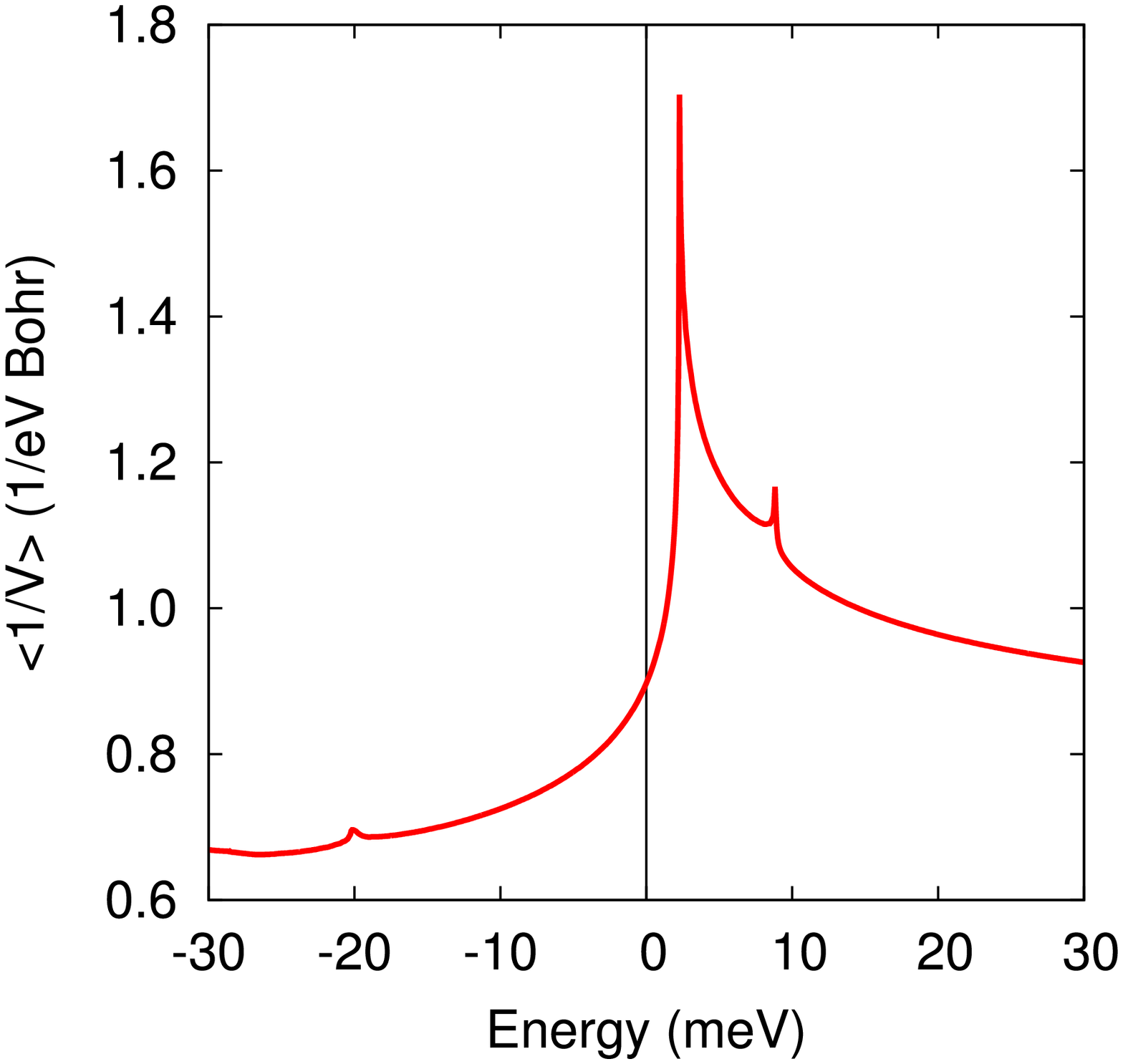}
\includegraphics[height=8.5cm,angle=-90]{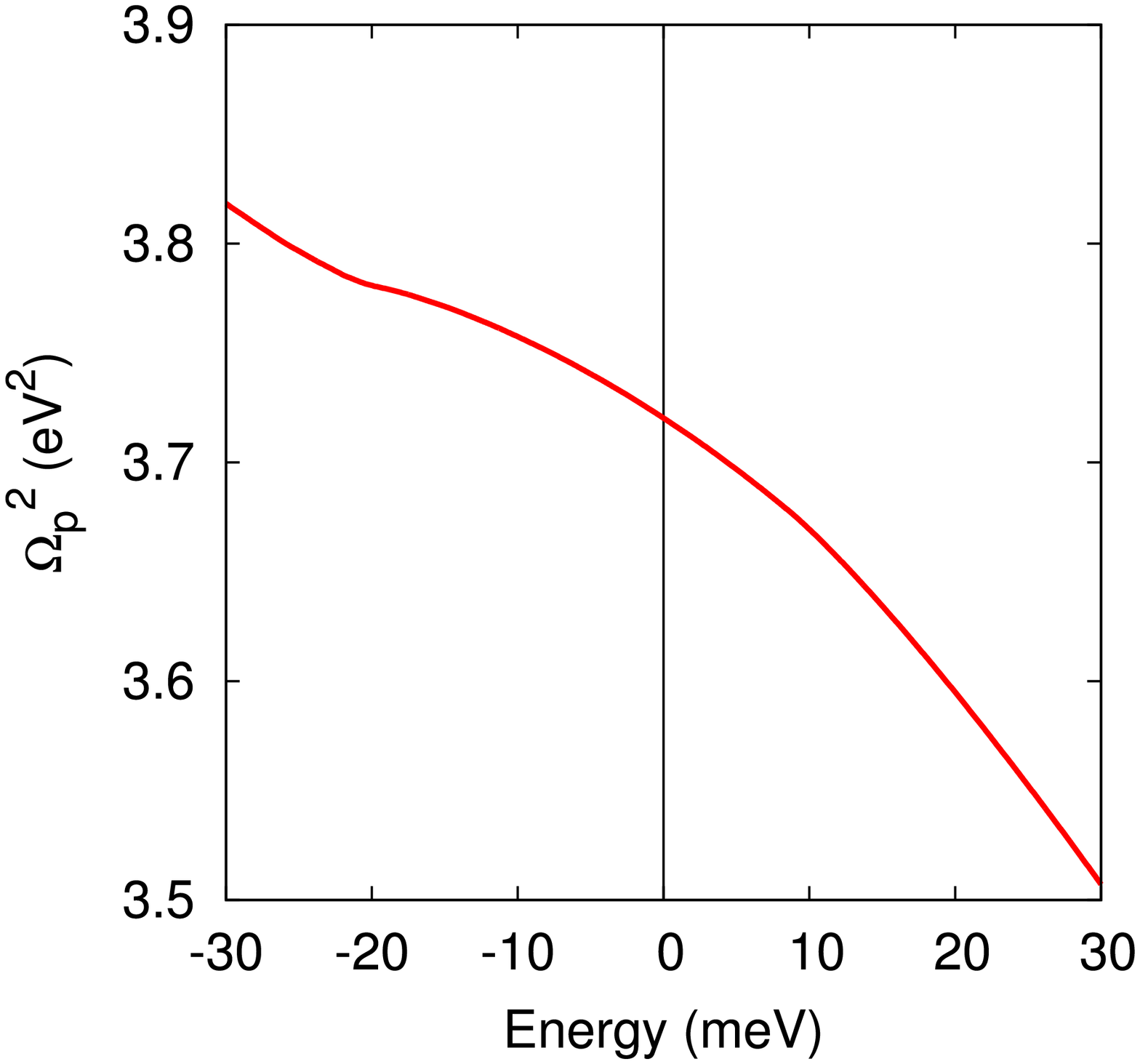}
\caption{(Color online) Top panel: $<\frac{1}{v(E)}>$ plotted versus 
energy, showing the square root divergence 
of the inverse moment of velocity near the Fermi energy. 
Unit conversion is: 1 eV Bohr = 8$\times 10^6$ cm/s.  Bottom panel:
the graph
of the second moment of velocity (with constants included to show it as
the square of the Drude plasma energy) is concave downward, which 
gives rise to the negative value of the Moriya $A$ parameter.  This
sign of $A$ is verified by the calculation of $\chi(q)$ at small $q$
(see text).
}
\label{moments}
\end{figure} 

\section{Analysis of Velocity Distribution and Susceptibility}
\subsection{Renormalization due to Spin Fluctuations}
Following the work of Larson, Mazin, and Singh\cite{MSA_Pd} for
Pd which builds on Moriya theory, we first attempted to identify the 
relevant band characteristics in order to evaluate the spin
fluctuation reduction of $\chi$ in TiBe$_2$.   For this, one begins with
the bare susceptibility in the small $q$ and small $\omega$
limit, given by
\begin{eqnarray}
\chi_{0}(\vec{q},\omega) = N(E_{F})\Large[1 - 
  A\large( \frac{qa}{2\pi}\large)^2 + i
  \frac{1}{2}<\frac{1}{v}>_F \frac{\omega}{q}\Large],
\label{moriya}
\end{eqnarray}
while the screened susceptibility using the RPA approximation is
given by
\begin{eqnarray}
\chi^{-1}(\vec{q},\omega) =  \chi_{0}^{-1}(\vec{q},\omega) - I.
\end{eqnarray}
The Moriya parameter $A = -1.8$, expressed in dimensionless form
here, and mean inverse 
Fermi velocity $<1/v>_F \equiv v^{-1}_F$
(the second Moriya parameter, discussed below) are derived from 
velocity moments and DOS of the band structure,
and like the density of states, they are greatly influenced by the Fermi
surface topology and its velocity spectrum.
Specifically, changes in topology which give rise to points of
zero velocity in the band structure near the Fermi surface become 
an important factor.
The mean inverse Fermi velocity which governs the imaginary part of
$\chi_{0}(\vec{q},\omega)$ is given by
\begin{eqnarray}
<\frac{1}{v(E)}> \equiv v^{-1}(E) = \sum_k 
        \frac{\delta(\varepsilon_k-E)}{|\vec v_k|} /
  \sum_k \delta(\varepsilon_k-E)
\end{eqnarray}
evaluated at E$_F$. The difference between $<v^{-1}>_F$ and 1/$<v>_F$ is
one measure of the velocity variation of the Fermi surface.
The bottom or top of a three-dimensional band (corresponding to
the appearance or vanishing of a Fermi surface) gives only a discontinuity
proportional to the square of the band mass.  At a saddle point,
such as the merging of the corners
of the pseudocube Fermi surfaces, 
$v^{-1}(E)$ undergoes a $1/\sqrt{E-E_{cr}}$ divergence because the associated
Fermi surface area does not vanish.   This ``van Hove 
singularity'' in $v^{-1}(E)$ is evident for the band edge 3 meV from E$_F$
in TiBe$_2$ in 
Fig. \ref{moments}. We calculated 
1/$v^{-1}_F=5$ 
$\times$ 10$^6$
cm/s for TiBe$_2$.

For cubic structures, the parameter $A$ in Eq. \ref{moriya} is given by
\begin{eqnarray}
A &=&\frac{1}{48\pi e^2}\Large(\frac{2\pi}{a}\Large)^2 
    \frac{d^2 \Omega_p^2(E_F)}{dE^2_F} 
  \\ \nonumber
\Omega_p^2(E_F)&=&\frac{4\pi e^2}{3} \sum_k {\vec v}_k^2 \delta(\varepsilon_k - E_F)
   \\ \nonumber
  &\equiv&\frac{4\pi e^2}{3} N(E_F) v_F^2.
\end{eqnarray}
Thus $A$ it is proportional to the second derivative of the
square of the Drude plasma energy $\Omega_p$ ({\it i.e.} $\hbar$ is
absorbed into $\Omega_p$, so $\Omega_p$ here
explicitly has energy units;  $k$ sums are understood to be normalized
over the zone).
The second moment of velocity is finite everywhere, but its
second derivative is not (for example, for free electrons this diverges
as the band edge). Derivatives have the unfortunate
property of amplifying noise in numerical evaluations. 
We have addressed the noise issue by using a
large number of k points in the numerical integration 
(360$\times$ 360$\times$ 360). By fitting
$\Omega_p(E)^2$ with a polynomial near the Fermi energy, we obtain the
above-mentioned value $A=-1.8$.
The Fermi velocity was calculated to be $v_F = 2.3$ eV bohr
= 1.8 $\times 10^7$ cm/s.

\subsection{{\rm q}-dependent Susceptibility}
\begin{figure}
\includegraphics[height=8.5cm,angle=-90]{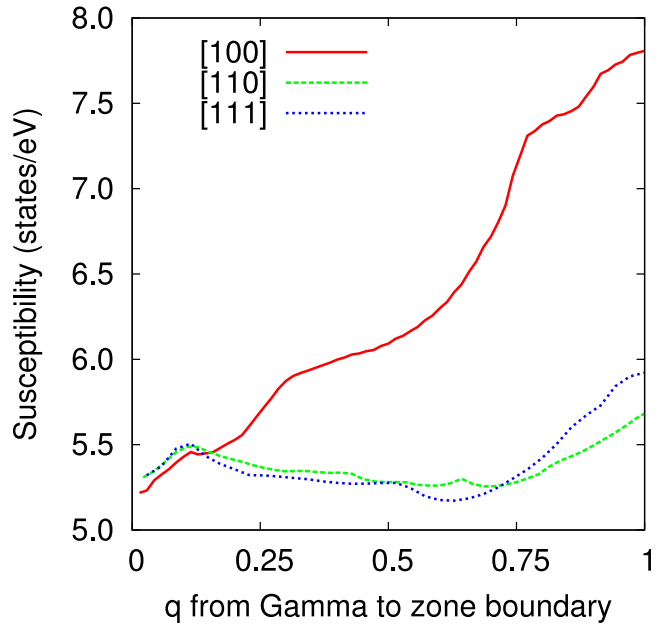}
\caption{(Color online) Intraband contribution to the real part
of $\chi(\vec{q})$. The increase at small $q$ confirms the sign of
Moriya $A$ coefficient (see text).  Although both [110] and [111]
directions have a maximum at the zone boundary, the peak along
[100] (X point of the zone) dominates the instability. 
}
\label{chi}
\end{figure}

The negative value of the $A$ parameter indicates, from Eq. \ref{moriya},
that the primary magnetic instability in TiBe$_2$ does {\it not} lie at
$q$=0 but rather at finite $q$, so it is  
more susceptible to AF instability (including possibly a spin spiral)
rather than ferromagnetic. The sign of $A$
has been verified independently by explicit calculation of the real part of 
$\chi(\vec{q})$, with results shown in Fig. \ref{chi}.   

The calculation of $\chi_{\alpha,\beta}(\vec{q})$ between bands
$\alpha$ and $\beta$ was performed by an isosurface slicing method.
The susceptibility can be written, after inserting a factor
$1 \equiv \int d\Delta ~\delta(\Delta
-\epsilon_{\beta,\vec{k}+\vec{q}}+\epsilon_{\alpha,\vec{k}})$, as
\begin{eqnarray}
\chi_{\alpha\beta}(\vec{q}) &=& 
\int d\Delta \frac{N^{\chi}_{\alpha\beta}(\Delta)}{\Delta},
\\ \nonumber
N^{\chi}_{\alpha\beta}(\Delta)&=&\sum_k \Large[f(\epsilon_{\alpha,\vec{k}}) -
f(\epsilon_{\beta,\vec{k}+\vec{q}})\Large]
\delta(\Delta
-\epsilon_{\beta,\vec{k}+\vec{q}}+\epsilon_{\alpha,\vec{k}}),
\end{eqnarray}
where $N^{\chi}_{\alpha\beta}(\Delta)$ is a susceptibility density calculated from the
isosurface defined by the Fermi functions and the energy $\delta$
function.  The Brillouin zone was divided into a $140\times 140\times 140$
grid of cubes.  Within each cube
the $\Delta$ integral is calculated as a discrete sum, using variable step
sizes in $\Delta$ corresponding to 1/30 of the maximum difference in
energies $\epsilon_{\beta,\vec{k}+\vec{q}}+\epsilon_{\alpha,\vec{k}}$
within the cube.

The susceptibility 
rises equally along all three symmetry directions (as required by cubic
symmetry), but only for $\hat q$ along the cubic axis does $\chi(\vec q)$
continue to increase strongly beyond the small-$q$ region.
The maximum of $\chi_o(\vec{q})$
occurs at the X point, where the intraband part has increased by nearly
50\% over its $q$=0 value.  In such cases where $q$=0 is not the
maximum, it is necessary to
apply the extension of weak ferromagnets to the AF case.\cite{weakAF}

The band-by-band contributions to $\chi_o(q)$ have been evaluated both
to verify the code and to identify the source of the important
contributions and structures.  The sphere FS gives rise to a Lindhard
type form with 2$k_F \approx \pi/a$ (but is not perfectly round).  The pillows 
lead to a cusp for $(q_x,0,0)$ for $q_x \approx 0.28 \pi/a$, and along
all three directions decreases for $q \geq \pi/a$.  For the
jungle gym and the pseudocube,$\chi$ increases by a factor of 
two at the zone boundary
along $(q_x,0,0)$, with much less variation in the other two directions.
The contributions to N(E$_F$) from each of the bands is: sphere, 1.4\%;
pillows, 7\%; jungle gym, 33\%; pseudocube, 58\%.

Away from $q$=0 the interband contributions to $\chi(q)$ contribute,
and it is known in other transition metals and their compounds 
that the $\vec q$-dependence
of matrix elements can be important.  We have calculated also the 
interband $\chi(\vec q)$ for several bands around the Fermi level, finding
that they contribute a broad maximum at intermediate $|q|$.  It seems
unlikely, however, that interband contributions will move the maximum
away from the X point.

Peaking of $\chi(\vec q)$ at the zone boundary implies a short
wavelength $\lambda = a$ AF instability (incipient, since no AF phase
is observed).  With the fcc lattice and two Ti atoms in the 
primitive cell, there several possibilities for the most
unstable mode, which will involve antialignment of spins or charge
density wave variation, but
also may involve noncollinear alignment of the spins.
We have tried to obtain a $q=0$ AF state within
LDA, with atomic moments antialigned on the bipartite Ti lattice, but
the moment vanished when this was tried.  We have not investigated
possible $\vec q$ = X point AF states.

\subsection{Temperature Dependence of Susceptibility}
The high narrow peak in the DOS near E$_F$ suggests an explanation of
the T-dependence of $\chi$ mentioned in the Introduction, or at least
part of it.  To understand what part arises from simple thermal
smearing, we have evaluated
\begin{eqnarray}
N(E,T) \equiv \int_{-\infty}^{\infty} 
  [-\frac{\partial f(E-\mu(T))}{\partial E}]
  N(E) dE,
\end{eqnarray}  
where the chemical potential $\mu(T)$ is adjusted at each temperature
to keep the number of electrons (occupied states) constant.  The 
result is shown as a series of curves for T ranging from zero to
300 K.  It is necessary to include the variation in $\mu$, and the value 
of $N(\mu(T),T)$ decreases by 8\%.

\begin{figure}
\includegraphics[height=8.5cm,angle=-90]{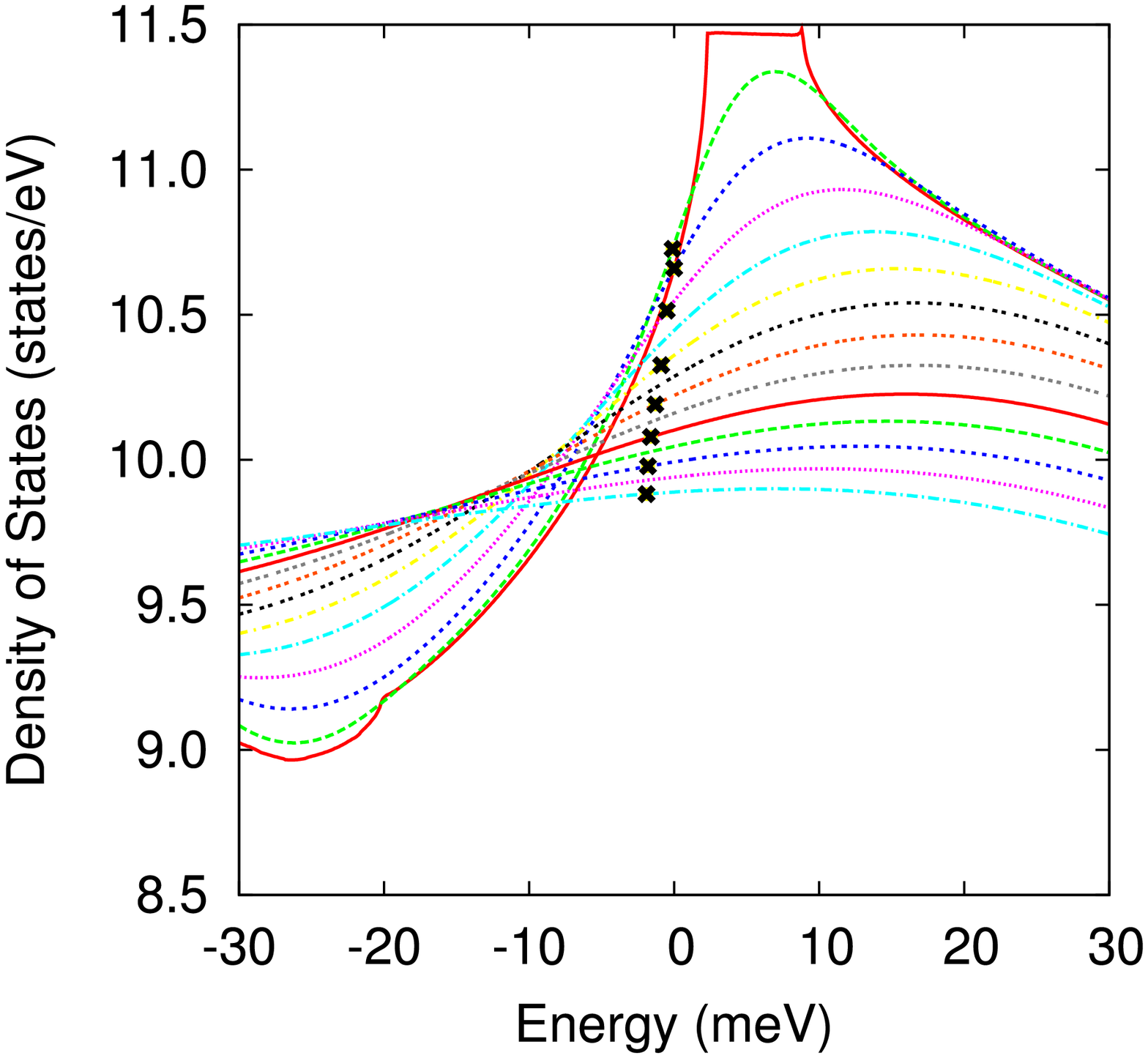}
\includegraphics[height=8.5cm,angle=-90]{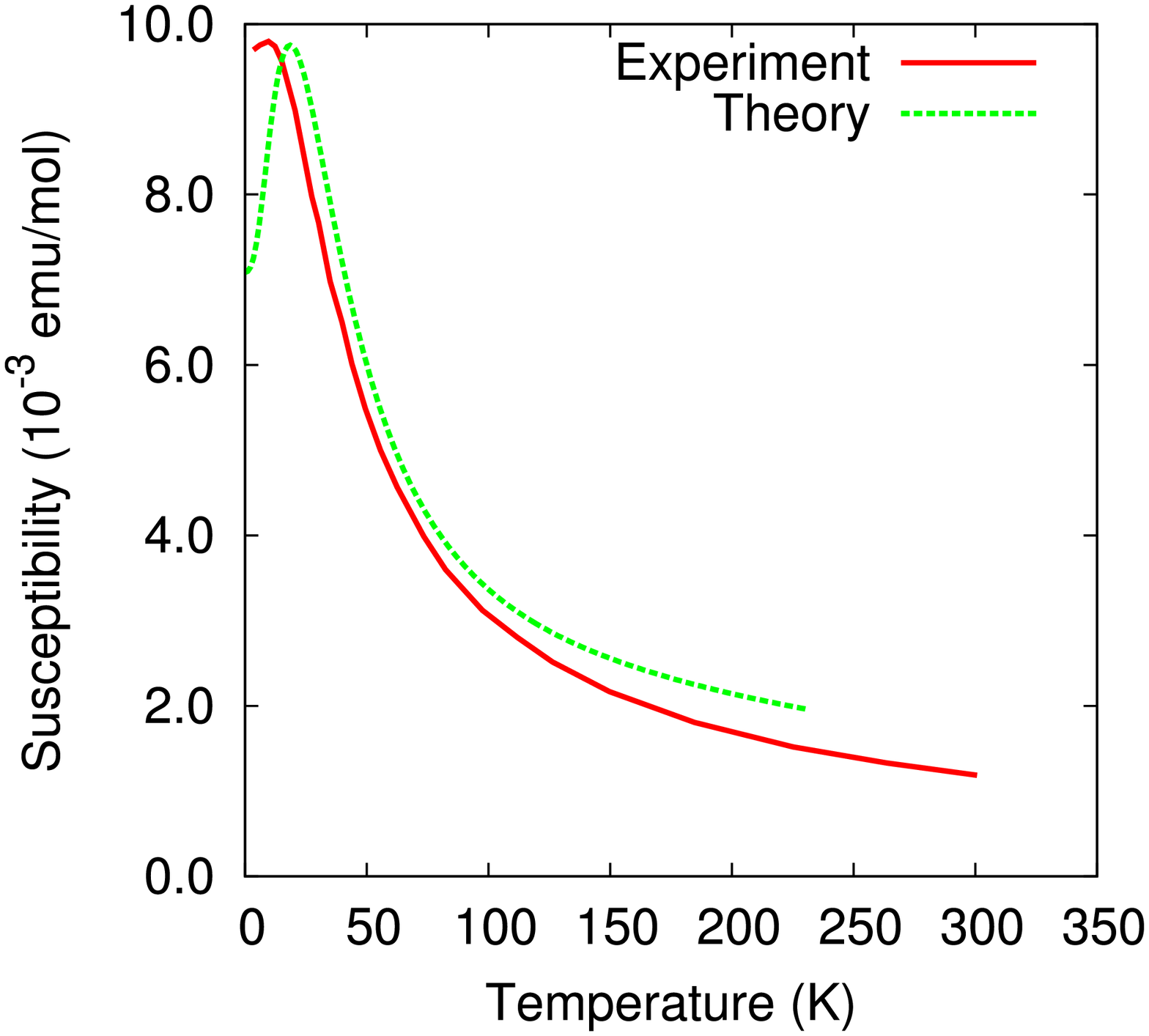}
\caption{(Color online) The upper graph shows how the density of states
near the Fermi energy changes from T=0 to T=300K. The lower
graph shows the experimental susceptibility\cite{gerhardt}
compared to theory. The Stoner I has been adjusted slightly from the calculated 
value to
match the susceptibility maximum.  
}
\label{N_T}
\end{figure} 

The resulting change in the physical, enhanced susceptibility is
given by
\begin{eqnarray}
\chi(T) = \frac{N(\mu(T);T)}{1-I N(\mu(T);T)}.
\end{eqnarray}
Adjusting I to reproduce the peak height (at 10 K, experimentally),
which requires I=0.183 eV ($S$=56 at the maximum of $N(\mu)$), 
the resulting enhanced $\chi(T)$ is
compared with the data in the lower panel of Fig. \ref{N_T}.  It is
evident that this simple temperature smearing accounts for much of
the observed temperature dependence.  Additional indirect 
temperature smearing will come from phonons and from electronic and
magnetic interactions as these excitations are increasingly excited
upon raising the temperature.  We conclude that TiBe$_2$ contains
no appreciable contribution to the susceptibility from local moments.

\subsection{Field Dependence of Susceptibility}
\begin{figure}
\includegraphics[height=8.5cm,angle=-90]{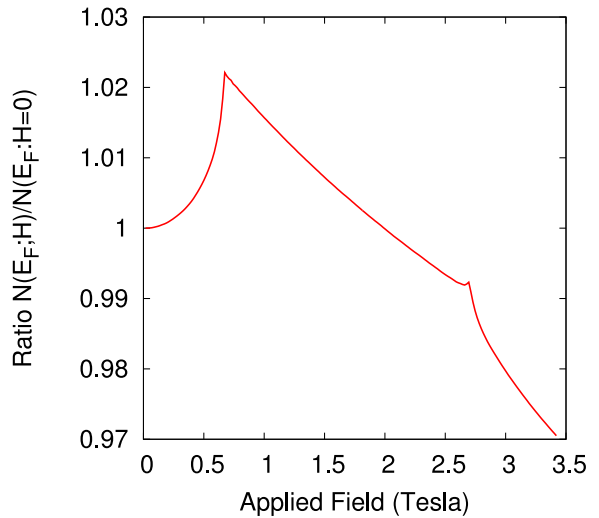}
\caption{(Color online) Magnetic field (H) dependence of the Fermi
level density of states for TiBe$_2$, referred to its H=0 value, as defined in
the text.  The initial increase with field indicates an increasing
instability towards ferromagnetic order.
}
\label{Ratio}
\end{figure}
For an energy-dependent DOS and a highly enhanced susceptibility,
a field-dependent susceptibility $\chi(H,T=0) \equiv \chi(H)$ is
expected.  In TiBe$_2$ a strong effect of this kind has been seen,
which can be characterized as field-driven ferromagnetism.  The 
differential susceptibility $\chi_d(H) = d{\cal M}(H)/dH$ where ${\cal M}$
is given by the difference in electron occupations $n_{\sigma}(H)$.  
A many-body treatment shows that the spin imbalance can be 
expressed\cite{wepPRL} in terms of the spin-dependent thermal
(energy E surface averaged) Green's function 
\begin{eqnarray}
G_{\sigma}(E,i\omega_n;H)&=&\frac{1}{i\omega_n -(E-\mu -\sigma \mu_B H)-
       \Sigma_{\sigma}(H)}, \\ \nonumber
n_{\sigma}(H)&=&\int dE N(E) T\sum_i 
    G_{\sigma}(E,i\omega_n;H)e^{i\omega_n \eta},
\end{eqnarray}
here $\omega_n$ is the fermionic Matsubara discrete energy variable
and $\eta$ is a positive infinitesimal.  The simplest form of 
(Stoner) self-energy $\Sigma_{\sigma} =\sigma K \mu_B H$ should be
appropriate ($1+K = S$).  

Taking the field derivative of ${\cal M}(H)$ but keeping H finite,
and using (in this approximation)
\begin{eqnarray}
G_{\sigma}(E,i\omega_n;H)=G_{\circ}(E-\sigma \mu_B (1+K)H,i\omega_n;H=0)
\end{eqnarray}
we obtain the result at zero temperature
\begin{eqnarray}
\chi_d(H) &=& \frac{d{\cal M}(H)}{d(\mu_B H)} \\ \nonumber
   &=&S \bigl[N(E_F -S \mu_B H) + N(E_F +S\mu_B H)\bigr].
\end{eqnarray}
This clearly reduces to the usual T=0 result at H=0.
A slightly better treatment would have also some H-dependence of
S due to the structure in $N(E)$ and the delicate situation here
that $I N(E_F)$ is approaching unity, but at this point we neglect
such details.

The result for the relative correction
\begin{eqnarray}
{\cal R} = \frac{N(E_F -S \mu_B H) + N(E_F +S\mu_B H)}
                {2N(E_F)}
\end{eqnarray}
is shown in Fig. \ref{Ratio}.  The effect on the ratio (thus on the
differential susceptibility) is clear, however even with the factor
of S=60 enhancement of the energy scale ($\mu_B H \rightarrow
S \mu_B H$) the peak occurs at a field one order of magnitude smaller
than seen in experiment.  This difference seems to indicate that
the field influence on the spin fluctuations dominates; however the
variation in N(E) will need to be accounted for in any quantitative
explanation.

\section{Summary}
The complex and sometimes confusing data on the enhanced paramagnet
were discussed in the Introduction.  It seems clear that magnetic
fluctuations will be required to understand the underlying
mechanisms.  Here we have presented a precise calculation and
analysis of the electronic structure, especially focusing on the
Fermi surfaces and velocity spectrum at and near the Fermi level
that underlies not only the single particle excitations but also
the spectrum of magnetic fluctuation in the itinerant limit, which
clearly seems to be the case in TiBe$_2$.

Our calculations have confirmed the sharp structure in the density
of states around the Fermi level that had been noted earlier,
and quantified the tiny energy scale that is involved: the Fermi
level lies in a region of steep DOS, just 3 meV from an abrupt
van Hove singularity.  This singularity is derived from a doubly
degenerate band at the L point of the zone.   We have shown how
to calculated the spectrum of velocities (speeds) over the 
Fermi surface, and find the spectrum to be peaked at (the low
value of) $10^7$ cm/s, with much of the weight below that value.
Moriya theory for weak ferromagnets requires, for the imaginary
part of the inverse susceptibility, the moment $<1/v_F>$; we have
illustrated that $1/v(E)$ diverges at the van Hove singularity
signaling possible problems with applying Moriya theory to
TiBe$_2$.  

Moriya theory for weak ferromagnets also requires the dimensionless
quantity $A \propto d^2 \Omega_p(E)/dE^2$ at the Fermi energy, 
where $\Omega_p$
is the conventional transport Drude energy.  We find that this
quantity is not positive, as it must be for an incipient 
ferromagnet; rather it is negative indicating the dominating (nearby)
magnetic instability is finite $q$: antiferromagnetic, spin wave,
spin spiral, etc.  Direct calculation of the generalized susceptibility
$\chi_o(q)$  confirms the sign of $A$, and reveals the dominant
instability to lie at the X point of the Brillouin zone, making
TiBe$_2$ an incipient antiferromagnet.

We have shown that the sharp structure in $N(E)$ has other consequences.
First, it leads to a T-dependent chemical potential.  Together with
the temperature broadening of $N(E)$ and the Stoner enhancement 
$S\approx 60$, this simple temperature broadening can account for most
if not all of the temperature dependence of the susceptibility,
which some investigators had interpreted as Curie-Weiss-like.  As a
result, the occurrence of local moments in TiBe$_2$ can be ruled
out.  Similarly, we have shown that this sharp structure in $N(E)$,
again together with the large Stoner enhancement, has a substantial
effect on the field-dependence of the differential susceptibility.
There is still the question of how much of the measured field dependence
is due to this induced exchange splitting, and how much is due to
the effect of the field on the magnetic fluctuations.     
  
Many of the results we have obtained here are strongly dependent on 
details of the band structure and the position of the Fermi level.
That these results reflect realistically the mechanisms underlying
the many fascinating observations obviously requires that the band
structure formalism is applicable in detail to such systems and that
the calculations are accurate.  Another requirement is that of high
sample quality, that the stoichiometry is precise and that defect
concentration must be very low (simple impurity broadening will
affect behavior).  These questions must be addressed in deciding
whether to press onward to a more complete and more challenging
explanation that includes effects of both magnetic fluctuations and
the energy dependence of the density of states. 

\section{Acknowledgments}
We acknowledge illuminating communication with Z. Fisk, I. I. Mazin, 
G. R. Stewart, and D. J. Singh about weak ferromagnetism and 
materials issues.
This work was supported by DOE grant DE-FG03-01ER45876.

\end{document}